\documentclass{ws-mpla}

\usepackage{amsmath,amssymb}
\usepackage{latexsym,graphicx,epsfig}

\def\lb{\linebreak[4]}
\newcommand{\be}{\begin{equation}}
\newcommand{\ee}{\end{equation}}
\newcommand{\bes}{\begin{subequations}}
\newcommand{\ees}{\end{subequations}}
\newcommand{\bear}{\begin{equation}\begin{array}}
\newcommand{\eear}[1]{\end{array}\label{#1}\end{equation}}

\newcommand{\fr}[2]{\dfrac{{ #1}}{{ #2}}}

\def\cl{\centerline}

\newcommand{\bu}{$\bullet$\ }

\begin{document}
\markboth{I.F. Ginzburg}
{Lessons from all logs summation in Yukawa theories}

%%%%%%%%%%%%%%%%%%%%% Publisher's Area please ignore %%%%%%%%%%%%%%
\catchline{}{}{}{}{}
%%%%%%%%%%%%%%%%%%%%%%%%%%%%%%%%%%%%%%%%%%%%%%%%%%%%%%%%%%%%%%%%%%%

\title{\bf LESSONS FROM ALL LOGS SUMMATION IN YUKAWA THEORIES
}
\author{I.~F.~GINZBURG}
\address{Sobolev Institute of Mathematics SB RAS, pr. ac. Koptyug, 4,
Novosibirsk 630090, Russia\\
ginzburg@math.nsc.ru\\
Novosibirk State University, ul. Pirogova, 2, Novosibirsk 630090,
Russia}

 \maketitle

\pub{Received 19.10.2009}{}

%%%%%%%%%%%%%%%%%%%%%%%%%%%%%%%%%%%%%%%%%%%%%%%%%%%%%%%%%%%%%%%%%%%%%
\begin{abstract}
Some features of old results in the total summation of all logarithmic contributions of all diagrams in Yukawa theory are presented. We discuss some lessons from this picture for the description of Pomeron, odderon, etc. in QCD.

\keywords{Summation of asymptotics; Yukawa theories; gauge theory.}
\end{abstract}
%%%%%%%%%%%%%%%%%%%%%%%%%%

\ccode{12.38 Cy, 11.55 y, 12.39 St}

%\maketitle

\section{Introduction}

The recipes for the asymptotics of the Feynman diagrams in
dependence on its topology  at arbitrary Lagrangian were obtained
in 60-th (see Refs.\cite{Efras,Eden} and references therein).
The reduction of problem to the special case of diffractive
processes\lb $AB\to CD$ at
 \be
s=(p_A+p_B)^2\approx -u=(p_A-p_D)^2\gg |t|=|(p_A-p_C)^2)|, m_i^2\,\label{diffr}
  \ee
in the   Yukawa theory (with pseudoscalar or scalar charged bosons $\phi_i$)
 \be
L=4\pi g\bar{\psi}\gamma^5\tau^i\psi\phi_i+(4\pi )^2\lambda(\phi_i^2)^2 \label{Yukawa}
 \ee
allows us to obtain very simple  recipes suitable for more
ambitious problems\cite{GinSas}.

Based on these results we summarized {\bf  all logarithmic
contributions of all diagrams} for scattering amplitude, assuming
finite charge renormalization\cite{sumBGES}. The final results
were  published in reviews\cite{GES}. Variations of results for
running coupling constants were almost evident based on approach
from\cite{GS}.

We used the Mellin transform of amplitude defined separately for
positive and negative signature
\bear{c}
f^\pm(s,t)\equiv \fr{f(s,t)\pm f(u,t)}{2}\,,\\[2mm] f^\pm(s,t)=\fr{i}{2}\int\limits_{i-\infty}^{+\infty} dj \left(\fr{u-s}{2}\right)^j
\xi^\pm(j)F^\pm(j,t);\quad
\xi^\pm(j)=\fr{1\pm e^{i\pi j}}{2\Gamma(j+1)\sin\pi j}\,
\eear{Mellin}
with  inverse transformation defined in the standard manner. The
asymptotical behavior in $s$ is determined by the most right
singularity of $F(j,t)$ in the complex $j$=plane. This point
$j=j_0$ is the same for each diagram for given $AB\to CD$ process
in the fixed theory (Yukawa, $\phi^3$, gauge theory).

For the contributions of separate diagrams we used Feynman type
$\alpha$-representation described e.g. in Ref.\cite{BSh} and
developed in Ref.\cite{Efras}.  We use  normalization of
amplitudes in which $d\sigma\propto |f|^2/s$ without additional
dimensional factors for all processes with both bosons and
fermions.

With suitable normalization for the amplitudes with fermions in
the Yukawa theory these leading singularities of separate
diagrams are formed from poles at $j=0$ as
\be
F_{diagr}^Y=\sum a_kj^{-k}\,.\label{Yupoled}
 \ee
It corresponds  $f_{diagr}(s)=\sum \tilde{a}_k\log^k s$.
The poles at $j=-1$ describe higher twist corrections.

In the gauge theory like QED or QCD  contributions of separate
diagrams for flavor singlet $t$-channel contain multiple poles in
the points $j=n>0$ with integer $n$ with gauge dependent
coefficients. However, contributions of different diagrams
strongly cancel each other so that the gauge invariant
contribution of  each order of perturbation theory is formed from
poles at $j=1$ as
\be
F_{pert,n}^G=\sum b_k(j-1)^{-k}\,.\label{gaugepoled}
 \ee
It corresponds  $f_{pert,n}(s)=s\sum \tilde{b}_k\log^k s$.
The poles at $j=0$ describe higher twist corrections.

In the $\phi^3$ theory we have $F_{diagr}^{\phi^3}=\sum
c_k(j+1)^{-k}$. It corresponds  $f_{diagr}(s)=\sum
\tilde{c}_k\log^k/ s$.

The mentioned cancellation of contributions of separate diagrams
don't allow us to apply methods developed for Yukawa theory in
the gauge theory. Nevertheless, we think that some features of
results obtained in Yukawa theory can have relation to the
description of complete picture in the gauge theory.

%%%%%%%%%%%%%%%%%%%%%%%%%%%%%%
\section{The amplitude with positive signature}
%%%%%%%%%%%%%%%%%%%%%%%%%%%%%%%%%

We discuss first the processes with vacuum number exchange in
$t$-channel ($A\to C$) and with positive signature.

%%%%%%%%%%%%%%%%%%%%%%%%%%%%%%%%%%
\subsection{LLA results}
%%%%%%%%%%%%%%%%%%%%%%%%%%%%%%

The simplest picture appears in scalar theory ($g=0$). In this
case the  LL contributions of perturbation theory are given by
simple ladder diagrams. In the leading logs approximation (LLA)
amplitude was found in Refs.\cite{Saw},\cite{GS} as
solution of quadratic equation
\bear{c}
v_{LL}=F_{LLA,g=0},\,\qquad
v_{LL}^2-jv_{LL}+r_{LL}=0\,,\qquad r_{LL}=240\lambda^2\;\Rightarrow\\[3mm]
\Rightarrow\;v_{LL}=\fr{1}{2}\left(j+ \sqrt{j^2-4r}\right)\,.
 \eear{LLASaw}
The LLA amplitude has fixed square-root branch point in the
$j$-plane at $j=2\sqrt{r}$, i.e. $f(s,t)\propto s^{2\sqrt{r}}(ln
s)^{-3/2}$.

Next, we consider fermions with Yukawa interaction. In this case
one must consider different two-particle states in the
$t$-channel. That are two-boson states ($M$) and
fermion-antifermion ($f\bar{f}$) states with projection to vector
($V$). (Other $f\bar{f}$ states don't contribute to the vacuum
number exchange amplitude.) In this case LLA result
\eqref{LLASaw} takes place with simple variation. Both $v_{LL}$
and $r_{LL}$ become matrices $2\times 2$ with components
($M,\,V$). In particular, in the case of flavor $SU(2)$ symmetry
 \be
 r=\begin{pmatrix} 240\lambda^2&\sqrt{6}g^2\\
                  \sqrt{6}g^2&3g^2.\\
                  \end{pmatrix}
                  \label{v1}
\ee

%%%%%%%%%%%%%%%%%%%%%%%%%%%%%%%%%%%
\subsection{Complete amplitude}
%%%%%%%%%%%%%%%%%%%%%%%%%%%%%%%%%%%%%%

The detail description of asymptotics of all diagram allows to summarize all logarithmic contributions of all diagrams, i.e. all pole contributions \eqref{Yupoled}.
\begin{figure}[ph]
\centerline{\psfig{file=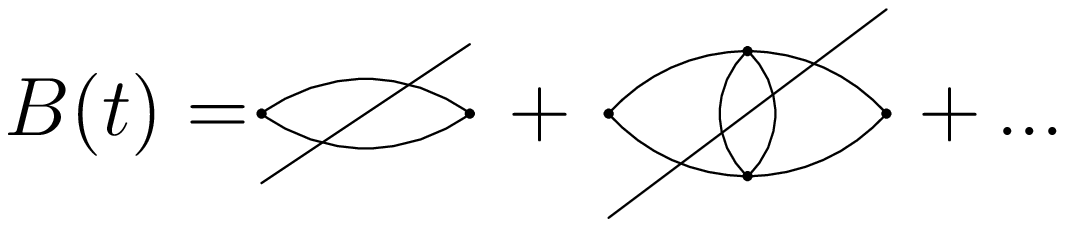,width=0.65\textwidth}}
%\hspace{0.05\textwidth}
\vspace{5mm}
\cl{\psfig{file=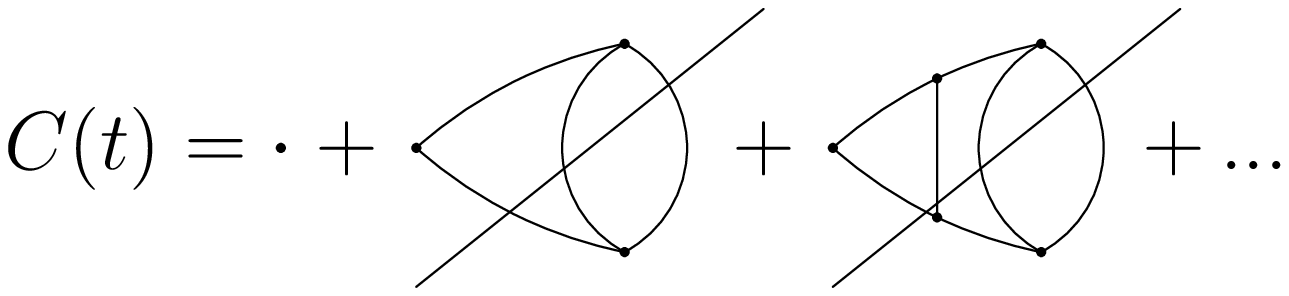,width=0.65\textwidth}}
\vspace*{8pt}
\caption{Diagrammatic representation for $B(t)$ and $C(t)$.\protect\label{figBC}}
\end{figure}

The results has form of $3\times 3$ matrix representation with
components ($M,\,V,\,T$):
 \be
F(j,t)=C^T_j(t)\left(\fr{1}{v^{-1}(j)-B_j(t)}\right)C_j(t)\,. \label{eqYuk}
 \ee
Here matrices $B(t)$ and $C(t)$ are described by diagrammatic
series of  Figs.~\ref{figBC}, where each term is specifically
regularized (oblique lines in pictures), it has no singularities
at $j>-1$, their $t$-dependence is most important fact. Matrix
$v(j)$ accumulates all poles at $j= 0$. It can be also presented
in the diagrammatic form where each term is independent on $t$
and particle masses. In the case of finite charge renormalization
$v(j)$ has fixed square-root branch point in the  $j$-plane. It
is obtained from quadratic equation system of form
\eqref{LLASaw}, where $r$ become $3\times 3$ matrix with
components ($M,\,V,\,T$); the LLA for this matrix is given by the
Eq.~\eqref{v1} (with $T$ components equal to 0). For the weak
coupling case the NLL terms of $v(j)$ are easily calculated and
fixed branch point position is shifted.

In the case of running coupling constant $v(j)$ has fixed
singularity in the $j$-plane, as before. To describe it,
following\cite{GS} one must analyze some subsidiary integral
equation  for $v(j)$.

It is easy to understand how this result will be changed if
vacuum expectation values of some products of field operators are
non-zero (vacuum condensate). In this case only equation for $B$
and $C$ are changed -- by adding of diagrams with insertion of
condensate blobs within diagrams shown in Figs.~\ref{figBC}.

Finally, this result gives representation of scattering
amplitude, containing both fixed points in $j$-plane and
Regge-poles (zeroes of denominator \eqref{eqYuk}). In our example
we cannot say even what of the contributions is dominant, the LLA
type fixed singularity in j-plane or Regge pole. Note that this
representation presents clear and unambiguous factorization --
all small distant contributions are contained in v(j) while large
distance contributions are contained in B(t) and C(t). 

We don't discuss in more details specific features of result in
the Yukawa theory.

\subsection{Lessons}\label{less1}

 \begin{enumerate}
\item
{\it Complete amplitude  has generally form, strongly different
from that of LLA.} In our example we cannot speak even what
contribution is dominant in real amplitude, LLA similar fixed
singularity in $j$-plane or Regge pole.

\item
{\it Improvements of LLA, similar to NLLA, NNLLA are often
misleading}. In our example they give corrected form for $v(j)$
and some general coefficient dependent on $t$ instead of correct
quite  different behavior.
\end{enumerate}

\section{Types of leading singularities of diagrams}

The studies of asymptotics of diagrams allow to distinguish  two
types of singularities of diagrams in $j$-plane (see
\cite{Efras,Eden}) in the case of finite masses
of all participating particles.

{\bu Small distance (end-point) singularities}. In the
$\alpha$-representation they originate from the region where for
some lines $\alpha_i\to 0$. These singularities appear in both
planar and non-planar diagrams.

{\bu Pinch singularities}. In the $\alpha$-representation they
originate from the region with generally non-zero $\alpha_i$ due
to some compensation between coefficients. These singularities
include effects of both  small and large distance effects. They
appear in non-planar diagrams only. In each theory the pinch
singularities in diagrams are responsible for the poles in
$j$-plane in the same point $j=j_0$ as for small distance
singularities.

For the considered class of theories we prove\cite{GinSas} {\bf
the theorem}, describing responsibility of different types of
singularities for asymptotics of amplitude for  summation of
poles  $(j-j_0)^{-n}$.\vspace{2mm}

{\centering{\begin{minipage}{0.85\textwidth}
\begin{itemize}

\item At  {\bf even $\pmb{j_0}$} (in particular, for ${j_0=0}$)\\
--- \ {\it the amplitude with positive signature} is described by  small
distance singularities only,\\
--- \ the pinch contributions together with small distance are
responsible for {\it the amplitude with negative signature}.

\item At {\bf odd $\pmb{j_0}$} (in particular, for ${j_0=\pm 1}$)\\
--- \ {\it the amplitude with negative signature} is described by only small
distance singularities,\\
--- \ the pinch contributions together with small distance are responsible
for {\it the amplitude with positive signature}.
\end{itemize}\end{minipage}}}\\%[3mm]

The summation of asymptotics with pinch singularities is very
complicated problem. Only the LL results were published here.
For the $\phi^3$  theory (with some preliminary estimates for
theories with higher $j_0$) it was done in Ref.\cite{Eden}.
These results were spread to QED in Ref.\cite{GrL}, the
studies of BFKL Pomeron\cite{BFKL} are in fact continuations of
this activity for pQCD.

%%%%%%%%%%%%%%%%%%%%%%%%%%%%%%
\section{Odderon/ Pomeron}
%%%%%%%%%%%%%%%%%%%%%%%%%%%%%%

Let us remind two points.

1. Pomeron is amplitude with positive signature, odderon  is
amplitude with negative signature.

2. In the theory $\phi^3$ we have $j_0=-1$; in the Yukawa theory
we have $j_0=0$; in the gauge theory (QED, QCD)  we have $j_0=1$.

Therefore:

In the Yukawa theory $j_0=0$, and the asymptotics of Pomeron
amplitude is determined by only small distance singularities of
diagrams. This very fact allowed us to summarize all logarithmic
asymptotics of all diagrams in the simple form \eqref{eqYuk}. The
odderon amplitude, which is determined by both small distance and
pinch singularities, was not calculated till now even in LLA.

In the $\phi^3$ theory $j_0=-1$, and the asymptotics of Pomeron
amplitude is determined by  both small distance and pinch
singularities of diagrams. It was found in LL as solution of some
integral equation\cite{Eden}. Vice versa,  the asymptotics of
odderon amplitude has much simpler structure, it is determined by
only small distance singularities of diagrams. The total
summation of all logarithmic contributions of all diagrams was
presented in Ref.\cite{GES}.

In the gauge theory (QED and QCD) $j_0=1$, and the asymptotics of
Pomeron amplitude is determined by  both small distance and pinch
singularities of diagrams. It is very complex problem. The LL
results were obtained in Ref.\cite{BFKL}. Recent NLL results
are note very clear for me due to reasons, discussed in
sect.~\ref{less1}. Vice versa,  the asymptotics of odderon
amplitude has much simpler structure, it is determined by only
small distance singularities of diagrams.The corresponding direct
analysis, starting from diagrams, was not done till now. The
representation\\ \cl{\it odderon = colored Pomeron + gluon} looks for us
misleading.

%%%%%%%%%%%%%%%%%%%%%%%%%%
\section*{Acknowledgment}
%%%%%%%%%%%%%%%%%%%%%%%%%%%%%

I discuss here mainly the results obtained in our collaboration
with Anatoly Efremov and our pupils. I recall this collaboration
as the most pleasant in my life. I am very thankful organizers of
ECT meeting "Recent Advances in Perturbative QCD and Hadronic
Physics" at July 20-25, 2009 for invitation. This paper was
supported by grants RFBR 08-02-00334-a and NSh-1027.2008.2.

\end{document}